\documentclass{jpp}
\usepackage{graphicx}
\usepackage{hyperref}
\usepackage[normalem]{ulem}
\usepackage[utf8]{inputenc}
\usepackage[T1]{fontenc}
\usepackage{amsmath}
\usepackage{comment}
\usepackage{color}

\newcommand{\bd}{\boldsymbol}
\newcommand{\ExB}{{\bd{E}\times\bd{B}}}
\newcommand{\exb}{{E\times B}}
\newcommand{\mc}{\mathcal}
\newcommand{\avg}[1]{\left\langle{#1}\right\rangle}
\newcommand{\pd}{\partial}
\newcommand{\etabar}{\bar{\eta}}
\newcommand{\rmb}{{\rm b}}
\newcommand{\rmd}{{\rm d}}
\newcommand{\rme}{{\rm e}}
\newcommand{\rmi}{{\rm i}}
\newcommand{\rmp}{{\rm p}}
\newcommand{\rmr}{{\rm r}}

\newcommand{\rmt}{{\rm t}}

\shorttitle{Collisionless zonal-flow dynamics in quasisymmetric stellarators}
\shortauthor{H. Zhu, Z. Lin, and A. Bhattacharjee}

\title{Collisionless zonal-flow dynamics in quasisymmetric stellarators}

\author{Hongxuan Zhu\aff{1}
  \corresp{\email{hongxuan@princeton.edu}},
  Z. Lin\aff{2},
 \and A. Bhattacharjee\aff{1}}

\affiliation{\aff{1}Department of Astrophysical Sciences, Princeton University, Princeton, NJ 08540
\aff{2}Department of Physics and Astronomy, University of California, Irvine, CA 92697}

\begin{document}

\maketitle

\begin{abstract}
The linear collisionless plasma response to a zonal-density perturbation in quasisymmetric stellarators is studied, including the  geodesic-acoustic-mode  oscillations and the  Rosenbluth--Hinton residual flow. While the geodesic-acoustic-mode oscillations in quasiaxisymmetric configurations are similar to tokamaks, they become non-existent in quasi-helically symmetric configurations when the effective safety factor in helical-angle coordinates is small. Compared with concentric circular tokamaks, the Rosenbluth--Hinton residual is also found to be multiplied by a geometric factor $\mc{C}$ that arises from the flux-surface averaged classical polarization. Using the near-axis-expansion framework, we derive an analytic expression for $\mc{C}$, which varies significantly among different configurations. These analytic results are compared with numerical simulation results from the global gyrokinetic particle-in-cell code GTC, and good agreement with the theoretical Rosenbluth--Hinton residual level is achieved when the quasisymmetry error is small enough.
\end{abstract}

\section{Introduction}
In axisymmetric magnetic confinement fusion devices, zonal flows are poloidal $\ExB$ flows which are toroidally symmetric but vary in the radial direction. Electrostatic zonal flows \citep{Lin98,Dimits00,Diamond05} (and their electromagnetic counterparts called ``zonal structures'' \citep{Zonca15,Dong19,Zocco23}) have been widely studied due to their role in regulating drift-wave turbulent transport.  Since the poloidal direction is not a symmetry direction in tokamaks, poloidal flows are expected to generate geodesic acoustic mode (GAM) 
oscillations \citep{Winsor68}, which are subject to collisionless Landau damping \citep{Conway21}. However, Rosenbluth and Hinton (RH) found that the zero-frequency branch of the zonal flow, where the divergence of the poloidal flow is balanced by the divergence of the parallel flow, do not experience collisionless Landau damping, so they can continuously grow while being driven by external source terms \citep{Rosenbluth98}. Supposing the source term is axisymmetric, the zero-frequency zonal-flow response is shielded by neoclassical polarization and reduced by a factor $1/(1+1.6q^2\epsilon^{-1/2})$ where $q$ is the safety factor and $\epsilon$ is the inverse aspect ratio. This factor is known as the RH residual-flow level, which is important because the residual zonal flow  can fully suppress turbulence near the linear instability threshold, which is known as the Dimits shift \citep{Dimits00}. The RH residual flow has also been widely simulated to test the validity and accuracy of gyrokinetic simulations \citep{Ye16,Moritaka19}.

Collisionless zonal-flow dynamics have also been studied in stellarators in the context of existing  experimental devices such as LHD, W7-X, HSX, and TJ-II \citep{Sugama06LHD,Mishchenko08,Helander11,Xanthopoulos11,Sanchez13,Monreal16,Monreal17,Smoniewski21,Nicolau21}.  It was found that after the initial GAM oscillations, zonal flows also experience slowly damped oscillations due to radially unconfined trapped particles. The RH level has been derived using both the gyrokinetic and the drift-kinetic formulation, which is written as a velocity-space integral. However, due to the complicated stellarator geometry, numerical calculation is usually required to evaluate the RH residual level.

In quasisymmetric (QS) stellarators \citep{Boozer83,Nuhrenberg88,Rodriguez20}, the magnitude of the magnetic field vector $\bd{B}$, which lies on flux surfaces, can be expressed as $|\bd{B}|=B(\psi,M\theta-N\varphi)$, where $\psi$ is
the flux surface label (defined as the toroidal magnetic flux divided by $2\pi$ in this paper), $\theta$ and $\varphi$ are the poloidal and toroidal angle in Boozer coordinates \citep{Boozer82}, and $M$ and $N$ are constant integers. This includes both quasi-axisymmetric (QA) devices where $M\neq 0$ and $N=0$, and quasi-helically (QH) symmetric devices where $M\neq 0$ and $N\neq 0$. (The quasi-poloidally symmetric devices with $M=0$ are not considered in this paper.) Since the drift-kinetic gyrocenter motion in QS stellarators is isomorphic to tokamaks in Boozer coorindates, the collisionless zonal-flow dynamics are expected to be also very similar. However, zonal flows in stellarators can still have geometry-specific properties. For example, a recent study pointed out that  due to the small effective safety factor, a high level of RH residual flow can be achieved in QH stellarators \citep{Plunk24} than tokamaks. With the progress in stellarator optimization, QS configurations with great accuracy have been designed \citep{Landreman22}, so the collisional neoclassical transport can be lowered to a level similar to tokamaks, and turbulent transport will be the dominant mechanism controlling confinement times \citep{Guttenfelder08,Beurskens21}. Since zonal flows often play a crucial role in regulating turbulent transport, we aim to make analytic progress in understanding zonal flows in QS stellarators, which is made easier due to the isomorphism in gyrocenter motion with tokamaks, when expressed in Boozer coordinates.

Here, we explore collisionelss zonal-flow dynamics in QS stellarators, including the GAM oscillation frequency and the  RH residual-flow level. The effects from gyroaveraging are not considered in this study, assuming the radial wavelength of zonal flows is  much larger than the ion gyroradius. We also assume the adiabatic-electron model since electrons have zero bounce-averaged radial drift in QS stellarators \citep{Mishchenko08}, but note that effects from kinetic electrons can be important for non-QS stellarators \citep{Monreal16,Nicolau21}. It is found that while the GAM oscillations in QA stellarators are similar to tokamaks, they become non-existent in QH stellarators when the effective safety factor in helical-angle coordinates is small. Compared to concentric circular tokamaks, the RH residual is also found to be multiplied by a geometric factor $\mc{C}$ that arises from the flux-surface averaged classical polarization $\avg{n_\rmi m_\rmi|\nabla\psi|^2/B^2}$. An analytical expression of $\mc{C}$ is obtained using the near-axis-expansion (NAE) framework \citep{Garren91a,Garren91b,Landreman19a,Landreman19b,Jorge20,Rodriguez22,Rodriguez23}, which varies significantly among different configurations. Note that similar modifications in the RH level has been found in tokamaks, which is mainly due to the flux-surface elongation \citep{Xiao06}. However, the elongation is limited by the vertical stability, so that typically $\mc{C}\lesssim 2.5$ \citep{Humphreys09,Lee15}. Here, a larger $\mc{C}$ (and the RH level) can be achieved for QA stellarators, provided that they are not subject to the vertical stability. Meanwhile, we found that $\mc{C}<1$ for QH stellarators, but the RH level is still enhanced due to the small effective safety factor \citep{Plunk24}. 

These analytic results are compared with numerical results from the global gyrokinetic particle-in-cell code GTC. We simulate zonal flows in 1st-order and 2nd-order NAE configurations, as well as the ``precise QA'' and ``precise QH'' configurations reported in \cite{Landreman22}. While the GAM physics is  reasonably predicted by the theory, we found that for the RH residual level, good agreement between analytical and numerical results is achieved only when the amplitude of QS-breaking magnetic-field component is small enough. As the next step of this research, we will study how the geometric factor $\mc{C}$ affects the nonlinear interactions between zonal flows and turbulence in QS stellarators. 

The rest of the paper is organized as follows. In section \ref{sec:theory}, we present our results on the RH level and the GAM frequency. In section \ref{sec:simulation}, we present numerical simulation results. Conclusions and discussions are given in section \ref{sec:conclusion}.
 
\section{Theory of collisionless zonal-flow dynamics}
\label{sec:theory}
\subsection{Calculation of Rosenbluth--Hinton residual flow in Boozer coordinates}
\label{sec:theory_RH}
Consider the time evolution of a zonal electrostatic potential $\Phi(\psi,t)$ and its associated radial electric field $E_r=-\pd_\psi\Phi|\nabla\psi|$. The RH residual flow can be understood from the conservation of toroidal angular momentum, where ``toroidal'' refers to the symmetric direction of the magnetic field \citep{Sengupta18}.  In an electrostatic gyrokinetic plasma, toroidal angular momentum consists of the $\ExB$-flow part $\mc{L}_{\exb}$ and the parallel-flow part $\mc{L_\parallel}$ \citep{Scott10,Brizard11,Stoltzfus-Dueck17,Zhu24}. The $\ExB$ part is defined as $\mc{L}_{\exb}=-\iota\avg{\bd{P}\cdot\nabla\psi}$ where $\iota$ is the rotational transform, $\avg{\dots}$ is the flux-surface average and the classical polarization $\bd{P}$ is obtained from $\nabla\cdot\bd{P}=e(Z_\rmi \delta n_\rmi-\delta n_\rme$). We have assumed a single gyrocenter ion species with mass $m_\rmi$, charge number $Z_\rmi$, density $n_\rmi= n_{\rmi 0}+\delta n_\rmi$, and temperature $T_\rmi=T_{\rm i0}$, while electrons are assumed adiabatic so their density perturbation can be written as $\delta n_\rme=n_{\rme 0}e(\Phi-\avg{\Phi})/T_\rme$ where $e$ is the elementary charge. Neglecting effects from gyroaveraging, we obtain $\bd{P}= -n_{\rmi 0} m_\rmi\nabla_\perp\Phi/eB^2$ from quasineutrality (see \eqref{eq:GAM_poisson} below), so that
\begin{equation}
\label{eq:RH_Lexb}
\mc{L}_\exb=\iota\Lambda_0\pd_\psi\Phi,\quad \Lambda_0=n_{\rmi0}m_\rmi\avg{|\nabla\psi|^2/B^2}.
\end{equation}
The parallel-flow part is defined as $\mc{L}_\parallel=\int d\bd{v}f_\rmi m_\rmi v_\parallel\hat{\bd{b}}\cdot\pd\bd{r}/\pd\varphi$, where $v_\parallel$ is the parallel velocity, $\hat{\bd{b}}=\bd{B}/B$, $f_\rmi(\bd{r},\bd{v},t)$ is the gyrocenter ion distribution, and we have neglected the electron contribution.  Assuming $\Phi$ evolves in time slowly compared to the trapped-ion motion, $\mc{L}_\parallel$ can be solved as the neoclassical plasma response to $E_r$ \citep{Rosenbluth98,Xiao06,Mishchenko08}. We obtain
\begin{equation}
\mc{L}_\parallel=\iota\Lambda_1\pd_\psi\Phi,
\end{equation}
where $\Lambda_1$ is given by \eqref{eq:RH_lambda} below. Assuming that a zonal density perturbation is applied to the plasma at $t=0$ such that $E_r$ is established without parallel flow; then, the plasma response will lead to GAM oscillations as well as the generation of parallel flow. For the linear zonal-flow dynamics where the perturbation is small, radial momentum transport (which is nonlinear) can be neglected, so that the toroidal angular momentum is conserved at each flux surface, $\Lambda_0\pd_\psi\Phi(\psi,t=0)=(\Lambda_1+\Lambda_0)\pd_\psi\Phi(\psi,t=\infty)$, from which we obtain the RH residual level as
\begin{equation}
\frac{E_r(t=\infty)}{E_r(t=0)}=\frac{1}{1+\Lambda_1/\Lambda_0}.
\end{equation}
Therefore, to evaluate the RH residual level in QS stellarator configurations, we need to quantitatively calculate $\Lambda_1$ and $\Lambda_0$.

A general expressions for $\Lambda_1$ has been derived by \citet{Mishchenko08} using the Boozer-coordinate representation, where the magnetic field can be written as
\begin{equation}
\bd{B}=\nabla\psi\times\nabla\theta+\iota\nabla\varphi\times\nabla\psi=G\nabla\varphi+I\nabla\theta+\delta\nabla\psi,
\end{equation}
where $G$, $I$, and $\delta$ are the covariant components of $\bd{B}$. To study both QA and QH configurations, we use a helical angle $\vartheta=\theta-N\varphi$ as the independent coordinate where $N$ is the toroidal mode number of $B$, so that the magnetic-field strength depends on $\vartheta$ but not $\varphi$. Then,
\begin{equation}   \bd{B}=\nabla\psi\times\nabla\vartheta+\iota_N\nabla\varphi\times\nabla\psi=G_N\nabla\varphi+I\nabla\vartheta+\delta\nabla\psi,
\end{equation}
where $\iota_N=\iota-N$ and $G_N=G+NI$. Therefore, for QH configurations with $|N|\gg|\iota|$, the effective rotational transform $|\iota_N|$ can be much larger than $|\iota|$ in helical-angle coordinates. We describe charged-particle gyrocenter orbits using their energy $\mc{E}$ and pitch-angle variable $\lambda=\mu/\mc{E}$ where $\mu$ is the magnetic moment. In QS stellarators, gyrocenter orbits include passing orbits $\mc{E}>\mu B_{\max}$ and trapped orbits where $\mc{E}\leq\mu B_{\max}$, and we can define the flux-surface average $\avg{\dots}$ and the bounce average $\overline{\dots}$ as
\begin{equation}
\label{eq:RH_flux_avg}
    \avg{f}=\frac{\int \rmd\vartheta\,\rmd\varphi\sqrt{g}f}{\int \rmd\vartheta\,\rmd\varphi\sqrt{g}},\quad     \overline{f}=\frac{\int\rmd\vartheta\,\rmd\varphi fB\sqrt{g}/v_\parallel}{\int\rmd\vartheta\,\rmd\varphi B\sqrt{g}/v_\parallel},\quad \sqrt{g}=\frac{G_N+\iota_N I}{B^2},
\end{equation}
where $v_\parallel=\pm\sqrt{2(\mc{E}-\mu B)/m_\rmi}$ is the parallel velocity and $\sqrt{g}=(\nabla\psi\times\nabla\vartheta\cdot\nabla\varphi)^{-1}$ is the Jacobian. For the bounce average, the integration is from $\vartheta=0$ to $\vartheta=2\pi$ for passing particles, and back and forth between bounce points for trapped particles.  Then, \cite{Mishchenko08} obtained
\begin{equation}
\label{eq:RH_lambda}
    \Lambda_1=4\pi\int \rmd v\rmd\lambda\frac{Z_\rmi^2e^2f_{\rmi 0}^2}{T_{\rmi 0}}v^3\left[\avg{\frac{B}{|v_\parallel|}\tilde{G}^2}-\avg{\frac{B}{|v_\parallel|}}^{-1}\avg{\frac{B}{|v_\parallel|}\tilde{G}}^2\right].
\end{equation}
Here, $v=\sqrt{2\mc{E}/m_\rmi}$, $\rho_\rmi=\sqrt{T_\rmi m_\rmi}/Z_\rmi e B$ is the gyroradius at thermal velocity, $f_{\rmi 0}$ is the Maxwellian distribution function, and the integration is only over the passing-orbit velocity space. Also, $\tilde{G}$ is the solution of
\begin{equation}
 v_\parallel\hat{\bd{b}}\cdot\nabla \tilde{G}=\bd{v}_\rmd\cdot\nabla\psi,\quad \bd{v}_\rmd=\rho_\parallel\nabla\times(v_\parallel\hat{\bd{b}}),\quad \rho_\parallel=m_\rmi v_\parallel/Z_\rmi eB.
\end{equation}
Note that we have simplified \eqref{eq:RH_lambda} compared to \cite{Mishchenko08} assuming that the bounce-averaged radial drift velocity is zero, $\overline{\bd{v}_\rmd\cdot\nabla\psi}=0$. 

We can further carry out the calculation of $\Lambda_1$ for QS magnetic fields where $B$ does not depend on $\varphi$, so that
\begin{equation}
v_\parallel\hat{\bd{b}}\cdot\nabla=\frac{\iota_N v_\parallel}{B\sqrt{g}}\pd_\vartheta,\quad \bd{v}_\rmd\cdot\nabla\psi=\frac{G_Nv_\parallel}{B\sqrt{g}}\pd_\vartheta\rho_\parallel,
\end{equation}
so that $\tilde{G}=G_N\rho_\parallel/\iota_N$.  Using the relation $\avg{B/|v_\parallel|}^{-1}=\overline{|v_\parallel|/B}$ for passing orbits, we obtain
\begin{equation}
\Lambda_1=4\pi G_N^2q_N^2\int \rmd\mc{E}\rmd\lambda \,\mc{E}\pd_{\mc{E}}f_{\rmi 0}\avg{\overline{\left(\frac{|v_\parallel|}{B}\right)}-\left(\frac{|v_\parallel|}{B}\right)},
\end{equation}
where $q_N=\iota_N^{-1}$ is the effective safety factor. Since the particle motion in QS stellarators is isomorphic to tokamaks in Boozer coordinates \citep{Boozer83}, the velocity-space integration can be calculated following the existing literature \citep{Rosenbluth98,Xiao06}. Writing the magnetic-field strength as $B=B_0[1+\epsilon\cos\vartheta+\mc{O}(\epsilon^2)]$ where $\epsilon\ll 1$ is a small parameter, $\Lambda_1$ is given by 
\begin{equation}
\Lambda_1=\frac{m_\rmi n_{\rmi 0} q_N^2G_N^2}{B_0^2}\left[1.6\epsilon^{3/2}+\mc{O}(\epsilon^2)\right].
\end{equation}

The evaluation of $\Lambda_0$, however, depends  on the geometry. In a large-aspect-ratio concentric circular tokamak with major radius $R_0$, $G=B_0R_0$ and $\psi\approx B_0r^2/2$ where $r=\epsilon R_0$ is the radius of the flux surface, we have $\Lambda_0=n_{\rmi 0} m_\rmi r^2$ and $\Lambda_1/\Lambda_0=1.6q^2\epsilon^{-1/2}+\mc{O}(\epsilon^0)$, which is the well-known RH result in tokamaks. In QS stellarators, however, $|\nabla\psi|$ varies significantly on a flux surface, so that the evaluation of $\Lambda_0$ is nontrivial and depends on the geometry. In the following, we use the NAE framework to derive an analytic expression of $\Lambda_0$. 

\subsection{Calculation of $\Lambda_0$ from the near-axis expansion theory}
\label{subsec:NAE}

The NAE framework provides a systematic approach to construct QS stellarator configurations. Given a prescribed set of parameters, QS configurations can be generated using NAE expansions up to 2nd order in $\epsilon$ (more details on the accuracy of the model can be found in section \ref{sec:simulation} below). However, since the RH residual is predicted accurately to the lowest order in $\epsilon$, we focus on parameters required to construct 1st-order QS configurations. Also, only vacuum fields are considered in the following because $I$ does affect $B$ to 1st order in $\epsilon$. Then, five quantities appear in the calculation of $\Lambda_0$ and the RH residual, including three from the axis shape $\bd{r}_0(\varphi)$, and another two quantities $\etabar$ and $\sigma(\varphi)$, which determine the flux-surface shaping and rotational transform. In particular, $\sigma(0)=0$ for 1st-order configurations that possess  stellarator symmetry (provided the axis also possesses such symmetry), and $\sigma(0)\neq 0$ for those which do not. Here, stellarator symmetry refers to a property of $\bd{B}$ that $(B_R,B_z,B_\phi)\to(-B_R,B_z,B_\phi)$ under $(R,z,\phi)\to(R,-z,-\phi)$ with respect to a reference point (chosen to be $z=\phi=0$) in cylindrical coordinates. Correspondingly, if $(R(\phi),z(\phi))$ is a field line then $(R(-\phi),-z(-\phi))$ is also a field line, including the axis \citep{Dewar98}.

Given a magnetic axis $\bd{r}_0(\varphi)$, we can calculate  its arc length $l(\varphi)=\int|\rmd\bd{r}_0/\rmd\varphi|\rmd\varphi$, curvature $\kappa(\varphi)$, and torsion $\tau(\varphi)$. We can also define orthonormal vectors along the axis, which are the tangent vector $\hat{\bd{t}}(\varphi)$, the normal vector $\hat{\bd{n}}(\varphi)$, and the binormal vector $\hat{\bd{\rmb}}(\varphi)$. These quantities are obtained through the following relations \citep{Mercier64,Landreman19b}:
\begin{equation}
\hat{\bd{t}}=\frac{\rmd \bd{r}_0}{\rmd l},\quad\kappa\hat{\bd{n}}=\frac{\rm d\hat{\bd{t}}}{\rmd l},\quad \hat{\bd{\rmb}}=\hat{\bd{t}}\times\hat{\bd{n}},\quad \tau\hat{\bd{n}}=-\frac{\rmd\hat{\bd{\rmb}}}{\rmd l},
\end{equation}
where $\rmd/\rmd l=(\rmd l/\rmd \varphi)^{-1}\rmd/\rmd\varphi$. Specifically, we obtain $\hat{\bd{t}}$ from the first equation (which by definition satisfies $|\hat{\bd{t}}|=1$), $\kappa$ and $\hat{\bd{n}}$ from the second equation assuming $\kappa>0$ and $|\hat{\bd{n}}|=1$, $\hat{\bd{\rmb}}$ from the third equation, and $\tau$ from the last equation. This procedure can be carried out when $\kappa$ does not vanish anywhere, which applies to the QA and QH configurations \citep{Landreman18}. For 1st-order vacuum QS configurations,  the magnetic fields are given by 
\begin{equation}
    \bd{B}=G_0\nabla\varphi,\quad  G_0=B_0R_0,\quad R_0=l(\varphi=2\pi)/2\pi,
\end{equation} 
where $B_0$ is the value of the magnetic field on the axis and $2\pi R_0$ measures the total length of the axis. Also, the Boozer toroidal angle $\varphi$ is defined such that $\rmd l/d\varphi$ is a constant, namely,
\begin{equation}
    \varphi= l/R_0.
\end{equation}
The corresponding equilibria are represented as
\begin{equation}
\label{eq:NAE_r}
    \bd{r}(\psi,\vartheta,\varphi)=\bd{r}_0(\varphi)+\epsilon\left[\frac{1}{\kappa}\cos{\vartheta}\hat{\bd{n}}(\varphi)+\frac{\kappa}{\etabar^2}(\sin{\vartheta}+\sigma\cos{\vartheta})\hat{\bd{\rmb}}(\varphi)\right]+\mc{O}(\epsilon^2).
\end{equation}
Here, $\epsilon=\etabar\sqrt{2\psi/B_0}$ where $\etabar$ is a constant in the model that describes the variation of $B$ along the flux surface; $\sigma=\sigma(\varphi)$ is the solution of the Riccati equation
\begin{equation}
\label{eq:NAE_sigma}
    \frac{\rmd\sigma}{\rmd\varphi}+(\iota_0-N)\left(1+\sigma^2+\frac{\etabar^4}{\kappa^4}\right)+\frac{2G_0\etabar^2\tau}{B_0\kappa^2}=0,
\end{equation}
where the on-axis rotational transform $\iota_0$ is found together with the solution $\sigma(\varphi)$ that satisfies the periodic boundary condition in $\varphi$. From \eqref{eq:NAE_r}, flux-surfaces with constant $\psi$ are rotating ellipses, which are characterized by their elongation $\tan \zeta$ and tilt angle $\Theta$ with respect to $\hat{\bd{n}}$. These two quantities can be obtained from \citep{Rodriguez23b} 
\begin{equation}
\label{eq:NAE_elongation}
    \sin (2\zeta)=\frac{2\etabar^2/\kappa^2}{1+\sigma^2+\etabar^4/\kappa^4},\quad\tan (2\Theta)=\frac{-2\sigma\etabar^2/\kappa^2}{1+\sigma^2-\etabar^4/\kappa^4}.
\end{equation}
(Note that $\Theta$ is a geometric poloidal angle measured in configuration space, which is not the same as $\vartheta$.) Therefore,  the flux-surface shape is determined by both $\etabar$ and $\sigma(\varphi)$, and $\sigma(0)=0$ for configurations that also possess stellarator symmetry.

Given a NAE configuration described above, we calculate $\Lambda_0$ as follows. Using the relation
\begin{equation}
\nabla\psi=\frac{1}{\sqrt{g}}\frac{\pd\bd{r}}{\pd\vartheta}\times\frac{\pd\bd{r}}{\pd\varphi},
\end{equation} 
and
\begin{equation}
    \frac{\pd\bd{r}}{\pd\vartheta}=-\frac{\epsilon}{\kappa}\sin\vartheta\hat{\bd{n}}+\frac{\epsilon\kappa}{\etabar^2}(\cos\vartheta-\sigma\sin\vartheta)\hat{\bd{\rmb}}+\mc{O}(\epsilon^2),\quad\frac{\pd\bd{r}}{\pd\varphi}=R_0\hat{\bd{t}}+\mc{O}(\epsilon),
\end{equation}
we have (to the lowest order in $\epsilon$) \citep{Jorge21}
\begin{equation}
\label{eq:NAE_gradpsi2}
    \frac{|\nabla\psi|^2}{B^2}=\frac{1}{B^2}\bigg|\frac{1}{\sqrt{g}}\frac{\pd\bd{r}}{\pd\vartheta
    }\times\frac{\pd\bd{r}}{\pd\varphi}\bigg|^2=\frac{\epsilon^2}{\etabar^2}\left[\frac{\etabar^2}{\kappa^2}\sin^2\vartheta+\frac{\kappa^2}{\etabar^2}(\cos\vartheta-\sigma\sin\vartheta)^2\right]
\end{equation}
and obtain
\begin{equation}
\Lambda_0=m_\rmi n_\rmi\avg{\frac{|\nabla\psi|^2}{B^2}}=\frac{m_\rmi n_\rmi\epsilon^2}{\etabar^2}\int\frac{1}{2}\left[\frac{\etabar^2}{\kappa^2}+\frac{\kappa^2}{\etabar^2}(1+\sigma^2) \right]\frac{\rmd\varphi}{2\pi}.
\end{equation}
The RH residual is then calculated as
\begin{equation}
\label{eq:NAE_RH}
 \frac{1}{1+\Lambda_1/\Lambda_0}=\frac{1}{1+1.6q_N^2\epsilon^{-1/2}/\mc{C}+\mc{O}(\epsilon^0)}.
\end{equation}
Compared to concentric circular tokamaks with the same $\epsilon$ and $q$, the RH residual in QS stellarators is modified by a geometric factor $\mc{C}$, which is given by
\begin{equation}
\label{eq:NAE_C}
    \mc{C}=\frac{1}{(\etabar R_0)^2}\int_0^{2\pi}\frac{1}{2}\left[\frac{\etabar^2}{\kappa^2}+\frac{\kappa^2}{\etabar^2}(1+\sigma^2) \right]\frac{\rmd\varphi}{2\pi}.
\end{equation}
The expected result $\mc{C}=1$ for concentric circular tokamaks can be recovered with $\etabar=\kappa=R_0^{-1}$ and $\sigma=0$. (Note that for tokamaks a nonzero on-axis current density should be included in the 1st-order NAE equations in order to have nonzero rotational transform.) For stellarator configurations with $\etabar\neq\kappa$ and $\sigma\neq 0$, we have $\etabar^2/\kappa^{2}+\kappa^2(1+\sigma^2)/\etabar^2>2\sqrt{1+\sigma^2}>2$ so that the integral is always larger than one, leading to possible enhancement of the RH residual.   The denominator $(\etabar R_0)^2$, however, depends on the configurations. Although $\etabar$ is a free parameter in the NAE theory, it is often chosen to maximize $\iota_0$ while approximately minimizing the flux-surface elongation at the same time \citep{Rodriguez23}. For the precise QA configuration studied in section~\ref{sec:simulation}, we found that $(\etabar R_0)^2<1$ and $\mc{C}>1$, leading to enhanced RH residual. For precise QH configuration, $(\etabar R_0)^2>1$ and $\mc{C}<1$, but the RH residual is still much larger  due to the small effective safety factor $|q_N|=|\iota_0-N|^{-1}$ \citep{Plunk24}. Also note that $\sigma(0)$ is often chosen to be zero so that the configuration possesses stellarator symmetry. From equation \eqref{eq:NAE_C}, it appears that non-stellarator symmetric configurations with nonzero $\sigma(0)$ could lead to larger $(1+\sigma^2)$ and hence larger RH residual. However, equation \eqref{eq:NAE_sigma} indicates that $(\iota_0-N)$ scales inversely with $(1+\sigma^2)$ at large $\sigma$ due to the periodic boundary condition in $\varphi$, so that the residual level does not necessarily increase with increasing $\sigma(0)$.

\subsection{Geodesic acoustic modes in quasisymmetric stellarators}
\label{subsec:GAM}
For numerical verification of the RH residual flow in a gyrokinetic code, one often initiates the simulation with a radially sinusoidal ion gyrocenter density perturbation and observe the time evolution of the corresponding radial electric field $E_r$. For these simulations, $E_r$ exhibits damped GAM oscillations at the beginning and reaches the stationary RH residual at the end. Since GAM oscillations are always present, it is also of interest to understand the GAM frequencies and damping rates. In tokamaks, the elongation is found to affect both the RH residual level \citep{Xiao06} and the GAM frequency \citep{Gao10}. Here, for QS stellarators, we expect the geometric factor $\mc{C}$ to play a similar role. In the drift-kinetic regime, a comprehensive analytic derivation of the GAM frequency in circular tokamak geometry has been given by \citet{Sugama06,Sugama08erratum,Gao08,Dorf13}. Here, we present an outline of the derivation from \cite{Sugama06}, which is slightly modified due to the QS stellarator geometry, as well as simplified assuming $\Phi=\avg{\Phi}$ for reasons discussed below. Under the radially local approximation, we write the ion gyrocenter distribution function as $f_{\rmi 0}+{\rm Re}(\delta f\rme^{\rmi k_\psi\psi})$  and the potential as ${\rm Re}(\Phi \rme^{\rmi k_\psi\psi})$, where $k_\psi$ is the wavenumber in $\psi$. Neglecting the gyroaveraging operator, the linearized gyrokinetic equation for ions is written as
\begin{equation}
\label{eq:GAM_GK}
    \left(\frac{\pd}{\pd t}+v_\parallel\hat{\bd{b}}\cdot\nabla+\rmi\omega_
    \rmd\right)\delta f=-(v_\parallel\hat{\bd{b}}\cdot\nabla+\rmi\omega_\rmd)f_{\rmi 0}\frac{e\Phi}{T_{\rmi 0}},
\end{equation}
where $\omega_\rmd=k_\psi\bd{v}_\rmd\cdot\nabla \psi$ is the drift frequency and $\hat{\bd{b}}=\bd{B}/B$. Note that here $\mu$ and $v_\parallel$ are treated as the independent velocity-space variables, namely, $v_\parallel$ no longer depends on spatial variables. This simplification is made assuming the GAM frequency $\sim v_{\rmt\rmi}/R_0$ is much larger than the ion transit frequency $\sim v_{\rmt\rmi}/qR_0$ \citep{Dorf13}. This assumption is justified for tokamaks with $q>1$, where the existing GAM theories have been developed and tested. For QS stellarators, this criterion will be replaced by $q_N/\sqrt{C}>1$ as discussed below.  For vacuum fields, $\sqrt{g}=G_N/B^2$, $G_N=G_0=B_0R_0$, and $B=B_0[1+\epsilon\cos\vartheta+\mc{O}(\epsilon^2)]$, we have
\begin{equation}
v_\parallel\hat{\bd{b}}\cdot\nabla\approx\frac{v_\parallel}{R_0q_N}\pd_\vartheta,\quad \omega_\rmd\approx\frac{v_\parallel}{R_0q_N}k_\psi\delta_\psi\sin\vartheta,
\end{equation}
where $\delta_\psi=\epsilon B_0R_0q_N(\rho_\parallel+\mu/Z_\rmi ev_\parallel)$ represents the neoclassical finite-orbit-width effects and the toroidal derivative $\pd_\varphi$ has been omitted for the zonal-flow dynamics. The potential $\Phi$ is solved from the long-wavelength limit of the gyrokinetic Poisson equation (quasineutrality condition):
\begin{equation}
\label{eq:GAM_poisson}
\nabla_\perp\cdot\left(\frac{n_{\rmi 0}m_\rmi}{eB^2}\nabla_\perp{\Phi}\right)=-(\delta \bar{n}_\rmi-\delta n_\rme),\quad \delta n_\rme=\frac{e(\Phi-\avg{\Phi})}{T_{\rme 0}}.
\end{equation}
Here, $\delta\bar{n}_\rmi$ is a gyroaveraged version of $\delta n_\rmi$ and will be approximated by the latter in the following. For concentric circular tokamaks, one can Fourier decompose in $\vartheta$, $\delta f=\sum_m\delta f_m\rme^{\rmi m\vartheta}$ and $\Phi=\sum_m\Phi_m\rme^{\rmi m\vartheta}$, and obtain the following results:
\begin{eqnarray}
\label{eq:GAM_circular_poisson}
   \frac{\delta n_0}{n_{\rmi 0}}=\avg{(k_\psi\rho_\psi)^2}\frac{e\Phi_0}{T_{\rmi 0}}, \quad \frac{\delta n_{m}}{n_{\rmi 0}}=\frac{e\Phi_m}{T_{\rme 0}} ~ {\rm for} ~ m\neq 0,
\end{eqnarray}
where $\delta n_m=\int \rmd^3 v\,\delta f_m$ and $\rho_\psi=\rho_\rmi|\nabla\psi|$ with $\rho_\rmi=\sqrt{m_\rmi T_{\rmi 0}}/Z_\rmi eB$ the ion gyroradius. For QS stellarators, however, $|\nabla\psi|^2/B^2$ varies significantly with $\vartheta$ and $\varphi$ (equation \eqref{eq:NAE_gradpsi2}), so that different poloidal and toroidal Fourier harmonics are coupled. While the solution for the zonal part $\avg{\Phi}$ is still given by $\Phi_0$ in \eqref{eq:GAM_circular_poisson} with only a small $\mc{O}(k_\psi^2\rho_\psi^2)$ correction, the solution for the non-zonal part $\Phi-\avg{\Phi}$ can be significantly different from $\Phi_{m\neq 0}$ in \eqref{eq:GAM_circular_poisson}, and solving them correctly can be a nontrivial task. For simplicity, we assume $\Phi=\avg{\Phi}$ and neglect the contribution from the non-zonal potential in the following. This is also consistent with the RH analysis where $\Phi=\avg{\Phi}$ has  been assumed for the calculation of $\mc{L}_{\exb}$ (equation \eqref{eq:RH_Lexb}), and can be achieved within the adiabatic-electron model assuming $T_{\rme 0}\ll T_{\rmi 0}$.

With the assumption that $\Phi=\avg{\Phi}$, the gyrokinetic equation \eqref{eq:GAM_GK} does not depend on $\varphi$, so that the Fourier components $\delta f_m$  are well defined. To solve $\delta f_m$  as a function of $t$, we apply Laplace transform in time, $\delta f_{m,\omega}=\int\rmd t\,\rme^{\rmi\omega t}\delta f_{m}$ and $\Phi_{0,\omega}=\int\rmd t\,\rme^{\rmi\omega t}\Phi_{0}$. The $m=0$ component of  \eqref{eq:GAM_GK} is
\begin{equation}
\label{eq:GAM_m=0}
-\rmi\omega\delta f_{0,\omega}-\delta f_0(t=0)
    =\frac{\rmi k_\psi\delta_\psi}{2R_0q_N}v_\parallel\left(\delta f_{-1,\omega}-\delta f_{1,\omega}\right). 
\end{equation}
To obtain $\delta f_{\pm1,\omega}$ as a function of $\Phi_{0,\omega}$, we write \eqref{eq:GAM_GK} as
\begin{equation}
     \left(\frac{\pd}{\pd t}+\frac{v_\parallel}{R_0q_N}\frac{\pd}{\pd\vartheta}\right)\left(\rme^{\rmi k_\psi\delta_\psi\cos\vartheta}\delta f\right)
     =
     -\frac{v_\parallel}{R_0q_N}\frac{\pd}{\pd\vartheta}\left(\rme^{\rmi k_\psi\delta_\psi\cos\vartheta}\frac{ef_{\rmi0}}{T_{\rmi 0}}\Phi\right).
\end{equation}
From the relation $\rme^{\rmi k_\psi\delta_\psi\cos\vartheta}=\sum_n\rmi^nJ_n(k_\psi\delta_\psi)\rme^{\rmi n\vartheta}$ where $J_n$ are the Bessel functions, we can solve for $\delta f_{ m,\omega}$ as  \citep{Sugama06}
\begin{equation}
\frac{\delta f_{m,\omega}}{f_{\rmi 0}}=\sum_{l,l'}\frac{\rmi^{l'-l}J_l(k_\psi\delta_\psi)J_{l'}(k_\psi\delta_\psi)}{\omega-(m+l)v_\parallel/R_0q_N}\left[\frac{(m+l)}{R_0q_N/v_\parallel}\frac{e\Phi_{m+l-l',\omega}}{T_{\rmi 0}}+\rmi\frac{\delta f_{m+l-l'}(t=0)}{f_{\rmi 0}}\right].
\end{equation}
The above expression can be simplified assuming $|k_\psi\delta_\psi|\ll 1$. Since we only consider the contribution from $\Phi_{0,\omega}$, we obtain $\delta  f_{1,\omega}$ as
\begin{equation}
\label{eq:GAM_f1}
    \frac{\delta f_{1,\omega}}{f_{\rmi 0}}=\left(\frac{k_\psi\delta_\psi}{2}\right)\frac{v_\parallel/R_0q_N}{\omega-v_\parallel/R_0q_N}\frac{e\Phi_{0,\omega}}{T_{\rmi 0}}
    +\left(\frac{k_\psi\delta_\psi}{2}\right)^3\frac{2(v_\parallel/R_0q_N)}{\omega-2(v_\parallel/R_0q_N)}\frac{e\Phi_{0,\omega}}{2T_{\rmi 0}}+\delta I_{1},
\end{equation}
and similarly for $\delta f_{-1,\omega}$. Here, higher-order (in $k_\psi\delta_\psi$) terms have been neglected, and $\delta I_1$ is from $\delta f_m(t=0)$. Note that the gyrokinetic Poisson equation \eqref{eq:GAM_circular_poisson} shows that $\delta f_0/f_{\rmi 0}$ is smaller than $e\Phi_0/T_{\rmi0}$ by a factor $(k_\psi\rho_\psi)^2$. Therefore, $\delta I_1$ can be neglected in \eqref{eq:GAM_f1} when the initial condition only consists of the $m=0$ component $\delta f_0(t=0)$, as is the common situation for numerical simulations.

Integrating \eqref{eq:GAM_m=0} over $(\mu,v_\parallel)$, together with \eqref{eq:GAM_circular_poisson} and \eqref{eq:GAM_f1}, one obtains
\begin{equation}
\Phi_{0,\omega}=(R_0q_N/v_{\rmt\rmi})\Phi_0(t=0)/K(\hat{\omega}).
\end{equation}
Here, $\hat{\omega}=\omega R_0q_N/v_{\rmt\rmi}$,  $v_{\rmt\rmi}=\sqrt{2T_{\rmi 0}/m_\rmi}$, and $K(\hat{\omega})$ is the GAM dispersion function:
\begin{equation}
\label{eq:GAM_dispersion}
    {K(\hat{\omega})}=-\rmi\hat{\omega}-\rmi\frac{q_N^2}{2\mc{C}}\bigg[2\hat{\omega}^3+3\hat{\omega}+(2\hat{\omega}^4+2\hat{\omega}^2+1)Z(\hat{\omega})+J_{\rm FOW}\bigg],
\end{equation}
where $Z(\hat{\omega})$ is the plasma dispersion function. Also, 
\begin{equation}
      J_{\rm FOW}=\rmi\frac{\sqrt{\pi}}{2}(k_\psi\delta_\psi)^2\rme^{-\hat{\omega}_\rmr^2/4}\left(\frac{\hat{\omega}_\rmr^6}{128}+\frac{\hat{\omega}_\rmr^4}{16}+\frac{3\hat{\omega}_\rmr^2}{8}+\frac{3}{2}+\frac{3}{\hat{\omega}_\rmr^2}\right)
\end{equation}
is from the resonance condition at $\omega=2v_\parallel/R_0q_N$, which was shown to significantly enhance the GAM damping rates. Compared to \citet{Sugama06,Sugama08erratum}, the geometric factor $\mc{C}$  appears in the ratio between $\delta_\psi^2$ and $\rho_\psi^2$:
\begin{equation}
    \frac{\avg{(k_\psi\delta_\psi)^2}}{\avg{(k_\psi\rho_\psi)^2}}\sim\frac{\epsilon^2R_0^2q_N^2}{(\avg{|\nabla\psi|/B)^2}}=\frac{q_N^2}{\mc{C}}.
\end{equation}
Therefore, compared to the tokamak results, here for QS stellarators we replace $q$ with $q_N/\sqrt{\mc{C}}$ except for the definition of $\hat{\omega}$. 

The evolution of $\Phi$ with $t$ is obtained through $\Phi(t)=\int\rmd\omega\,\rme^{-\rmi\omega t}\Phi_{0,\omega}/2\pi$ where the integration is from $-\infty+\rmi\gamma_0$ to $+\infty+\rmi\gamma_0$ with any positive real $\gamma_0$. Letting $\omega=\omega_\rmr+\rmi\gamma$ and $\hat{\omega}=\hat{\omega}_\rmr+\rmi\hat{\gamma}$, the GAM frequencies are found from $K(\omega)=0$ in the lower complex plane. Analytic results can be obtained using the asymptotic expansion of $Z(\hat{\omega})$ assuming $|\hat{\omega}|\gg 1$ and $|\gamma|\ll|\omega_\rmr|$, resulting in \citep{Sugama06,Sugama08erratum}
\begin{multline}
\label{eq:GAM_frequency}
\qquad\qquad\quad\omega_\rmr=\frac{\sqrt{7}}{2}\frac{q_N}{\sqrt{\mc{C}}}\left(\frac{v_{\rmt\rmi}}{R_0q_N}\right)\left(1+\frac{46}{49q_N^2/\mc{C}}\right)^{1/2},
\\
   \gamma=-\frac{\sqrt{\pi}}{2}\frac{q_N^2}{\mc{C}}\left(\frac{v_{\rmt\rmi}}{R_0|q_N|}\right)\left(1+\frac{46}{49q_N^2/{\mc{C}}}\right)^{-1}\bigg[\rme^{-\hat{\omega}_\rmr^2}\left(\hat{\omega}_\rmr^4+\hat{\omega}_\rmr^2\right)
    \\
    +\frac{1}{4}\left(k_\psi\delta_\psi\right)^2\rme^{-\hat{\omega}_\rmr^2/4}\left(\frac{\hat{\omega}_\rmr^6}{128}+\frac{1}{16}\hat{\omega}_\rmr^4+\frac{3}{8}\hat{\omega}_\rmr^2\right)\bigg].
\end{multline}
Therefore, $\omega_\rmr q_NR_0/v_{\rmt\rmi}\sim q_N/\sqrt{C}$ and $|\gamma/\omega_\rmr|\sim (q_N/\sqrt{C})\exp(-q_N^2/\mc{C})$, so that GAM oscillations are expected to be heavily damped in QH configurations with small $q_N^2/\mc{C}$. Note, however, that the ratio between the GAM frequency and the transit frequency is $q_N/\sqrt{C}$, which should be larger than one in order for the above GAM theory to be valid. While such criterion is generally satisfied for QA configurations studied in section~\ref{sec:simulation} below, it is not satisfied for QH configurations where $q_N/\sqrt{C}<1$, so that the above GAM theory may not quantitatively describe the heavy GAM damping in QH configurations.

\subsection{Application beyond the near-axis expansion}
\label{subsection:beyond_NAE}
Although the NAE description allowed us to derive an analytical expression of $\mc{C}$ \eqref{eq:NAE_C}, it is not required for the theoretical description of the RH residual and the GAM oscillations. Here, we examine the assumptions behind these theories and their validity for general QS stellarators beyond the NAE description.

The RH residual flow is a result of the toroidal angular momentum conservation, which is a general result in QS configurations, and the expressions \eqref{eq:RH_Lexb} and \eqref{eq:RH_lambda} for $\Lambda_0$ and $\Lambda_1$ are also general. Therefore, as long as the magnetic-field strength satisfies 
\begin{equation}
\label{eq:NAE_B}
    B=B_0[1+\epsilon\cos\vartheta+\mc{O}(\epsilon^2)], 
\end{equation}
we will have $\Lambda_0\propto \epsilon^2$ and  $\Lambda_1\propto 1.6q_N^2\epsilon^{3/2}$, and then the RH residual can still be written as $(1+1.6q_N^2\epsilon^{-1/2}/\mc{C})$ with a small parameter $\epsilon$ and a factor $\mc{C}$. While  $\mc{C}$ can be estimated from the axis shape using the NAE result \eqref{eq:NAE_C}, it can also be more accurately calculated from direct numerical evaluation of $\Lambda_0$. Suppose the relation \eqref{eq:NAE_B}
holds and $\mc{C}$ is obtained from either the NAE or direct numerical evaluation, the theory of GAM oscillations in section \ref{subsec:GAM} can also be carried out without assuming the NAE.

The relation \eqref{eq:NAE_B} holds for any QS stellarators near the axis where the NAE description is valid, where $\epsilon=\etabar r$ is proportional to the inverse aspect ratio and characterizes the variation of $B$ along field lines. As shown in section \ref{sec:simulation_precise}, this relation also holds very well for the precise QA and precise QH configurations, even if they are not obtained from the NAE approach. In fact, a recent work has shown that a large class of QS magnetic fields can described by the cnoidal solutions of the Korteweg-de Vries (KdV) equation, which are dominated by the $\cos\vartheta$ component even far away from the axis \citep{Sengupta23}. Therefore, we expect our theory of the collisionless zonal-flow dynamics to be applicable to a large class of QS stellarators beyond the NAE.
\section{Numerical simulations}
\label{sec:simulation}
\subsection{Simulation setup}
We use the global gyrokinetic particle-in-cell code GTC (gyrokinetic toroidal code\footnote{\url{https://sun.ps.uci.edu/gtc}}) to simulate collisionless zonal-flow dynamics. The code utilizes global field-aligned mesh in Boozer coordinates and has been verified for the simulation of microturbulence and zonal flows in the stellarator geometry \citep{Wang20,Fu21,Nicolau21,Singh23}. We choose a global code because for the non-axisymmetric stellarator geometry different radially local flux tubes could lead to different results, whereas a global code provides a simpler and more sharply defined setup for studying zonal flows. Note that previous studies also showed that flux-tube simulations give reasonable approximations to the global simulation results of the RH residual when the parallel extent of the flux tube is sufficiently long, but the flux-tube length required for convergence is configuration-dependent, for example, 4 poloidal turns for HSX \citep{Smoniewski21}, 2 poloidal turns for LHD, and at least 6 poloidal turns for W7-X \citep{Sanchez21}. 

We use single-species deuterium ions with $m_\rmi=2m_\rmp$, $Z_\rmi=1$, and uniform $n_{\rmi 0}$ and $T_{\rmi 0}$. At $t=0$, we choose a radial location $\psi_0$ and apply a radially sinusoidal perturbation in the ion weights in a narrow range $\psi\in[\psi_0-\Delta\psi/2,\psi_0+\Delta\psi/2]$ so that
\begin{equation}
    \frac{\delta f(t=0)}{f_{\rmi0}}=-w\sin\left(2\pi\frac{\psi-\psi_0}{\Delta\psi}\right),\quad w\ll 1.
\end{equation}
In other words, we apply a zonal-density perturbation with wavenumber $k_\psi=2\pi/\Delta\psi$ at the flux surface $\psi_0$, similar to the flux-tube simulations. We can apply the perturbation at different radial locations with varying $\psi_0$ and study the dependence of the RH residual on $\epsilon=\etabar r$ with $r=\sqrt{2\psi_0/B_0}$. We also choose $\Delta\psi=0.2\sqrt{\psi_0\psi_a}$ where $\psi_a$ is the value of $\psi$ at the outermost flux surface of the equilibrium, so that the zonal-flow wavelength $\Delta r\approx 0.1a$ is always 1/10 of the minor radius at the boundary $a=\sqrt{2\psi_a/B_0}$, and $k_\psi\rho_\psi$ and $k_\psi\delta_\psi$ (and hence the GAM frequencies) become independent from $\psi_0$. For each configuration, we choose 8 different values of $\psi_0$ corresponding to $r=0.2a, 0.3a$, ..., $0.9a$, which are evenly spaced and away from the inner and outer radial boundary $\psi_{\rm in}=0.01\psi_a$ and $\psi_{\rm out}=\psi_a$ used in the simulations. We note that as $r$ decreases the zonal-flow wavelength $\Delta r$ becomes comparable to $r$, so that $\epsilon$ becomes less well defined and the simulation results are expected to deviate from the  theory. For this reason, the radial location $r=0.1 a$ is not included,  even though it is still away from the inner boundary. At $t>0$, the ion weights are evolved from the delta-$f$ gyrokinetic equation
\begin{equation}
    (\hat{L}_0+\delta \hat{L})\delta  f=-\delta \hat{L}f_{\rmi 0},
\end{equation}
with
\begin{eqnarray}
    \hat{L}_{0}=\frac{\pd}{\pd t}+(v_\parallel\hat{\bd{b}}+\bd{v}_\rmd)\cdot\nabla-\frac{\mu\bd{B}^*\cdot\nabla B}{m_\rmi B}\frac{\pd}{\pd v_\parallel},~\delta \hat{L}=\bd{v}_E\cdot\nabla-\frac{Z_\rmi e \bd{B}^*\cdot\nabla\hat{J}_0\Phi}{m_\rmi B}\frac{\pd}{\pd v_\parallel}.
\end{eqnarray}
Here, $\bd{v}_E=\hat{\bd{b}}\times\nabla(\hat{J}_0\Phi)/B$ is the $\ExB$-drift velocity, $\hat{J}_0$ denotes gyroaverage on $\Phi$, $\bd{B}^*=\bd{B}(1+\rho_\parallel\nabla\times\hat{\bd{b}})$, and $f_{\rmi 0}$ is chosen to be Maxwellian. With the assumption $T_{\rme 0}\ll T_{\rmi 0}$, the potential $\Phi=\avg{\Phi}$ is obtained from the gyrokinetic Poisson equation \eqref{eq:GAM_poisson}.

In the following, we present simulation results for several 1st-order and 2nd-order vacuum QA and QH configurations obtained from the NAE approach \citep{Landreman19a,Landreman19b}, as well as the ``precise QA'' and ``precise QH'' configurations obtained from global optimization \citep{Landreman22}. These configurations are generated by VMEC\footnote{\url{https://princetonuniversity.github.io/STELLOPT/VMEC.html}}.  For the NAE configurations, the VMEC input files are generated by pyQsc\footnote{\url{https://landreman.github.io/pyQSC}}, which prescribes their fixed outermost flux surfaces at $a=\sqrt{2\psi_a/B_0}=0.1{\rm m}$ with $B_0= 1{\rm T}$. In other words, while their boundary are described by the NAE, these VMEC equilibria are still global and are not identical to the NAE inside the boundary \citep{Landreman19b}. For the precise QA and precise QH configurations, the corresponding VMEC equilibria are readily available from \citet{Landreman21data}, and the outermost flux surfaces correspond to $a=0.16{\rm m}$ and $a=0.11{\rm m}$, respectively. With the VMEC equilibria, the geometry and the magnetic fields are then converted to Boozer coordinates using BOOZ\_XFORM\footnote{\url{https://hiddensymmetries.github.io/booz_xform}}, which are used for the GTC simulations. Several geometric parameters of these configurations are summarized in table \ref{table:summary}, and all these configurations possess stellarator symmetry. For the numerical details, we choose $n_{\rmi 0}=10^{19}{\rm m}^{-3}$ in our simulations, which does not enter our results on the RH residuals and GAM frequencies. The choice of $T_{\rmi 0}$, however, requires further justification. The RH analysis assumed a small but finite $|k_\psi\delta_\psi|\sim|k_\psi\rho_\psi q_N/\sqrt{\mc{C}}|$, so that $T_{\rmi 0}$ cannot be too large. Since the stellarator configurations presented here have relatively small radius $r\approx 0.1{\rm m}$ and weak magnetic field $B_0\approx 1{\rm T}$, we choose $T_{\rmi 0}=1{\rm eV}$ for the QA configurations and $T_{\rmi 0}=5{\rm eV}$ for the QH configurations, which correspond to $|k_\psi\delta_\psi|\approx 0.15$ for the precise QA and precise QH configurations in section~\ref{sec:simulation_precise} below.  The mesh grids have a radial resolution of $a/200\approx 0.5{\rm mm}$ (20 grids per zonal-flow wavelength) and a poloidal resolution of $1{\rm mm}$ (about $5\rho_\rmi$). In the toroidal direction, we simulate one field period of the configurations with $N_p(N_e+1)$ planes. Here, $N_p$ planes are used where $\delta n_\rmi$ is calculated for solving $\Phi$, and an additional $N_e$ planes are inserted between each neighboring two of the $N_p$ planes where magnetic fields are interpolated for pushing particles \citep{Wang20}. We use $N_e=2$, and note that GTC prefers $N_p(N_e+1)+1$ to be even for the periodic cubic spline, so we choose $N_p=15$ for the QA configurations, and $N_p=31$ for the QH configurations. Approximately 100 marker particles per mesh node are used, and the simulation time step is $0.02 R_0/v_{\rmt\rmi}$. The simulation results are well converged for these choices of parameters.

\begin{table}
  \begin{center}
\def~{\hphantom{0}}
  \begin{tabular}{lccccccccccc}
      Configurations & $N_{\rm fp}$ & $N$ & $R_0$  & $q_N$ & $\etabar$ & $\mc{C}$ & RH residual & $\omega_\rmr^{\rm ana}$& $\omega_\rmr^{\rm num}$&$\gamma^{\rm ana}$&$\gamma^{\rm num}$ 
      \\[3pt]
       1st-order QA, a & 3&  $\,$0 & 1.02 & -2.92 & 0.60 & 5.11& 0.11 & 0.82 & 0.79 &-0.09 & -0.11
      \\
      1st-order QA, b & 3&  $\,$0 & 1.02 & -2.56 & 0.70 & 3.15& 0.09 & 0.96 & 0.96 & -0.06& -0.06
      \\
      1st-order QA, c & 3& $\,$0 & 1.02 & -2.44 & 0.80 & 2.18& 0.07 &1.09 & 1.10 & -0.05& -0.05
      \\
      2nd-order QA & 2 &  $\,$0 & 1.06 & $\,$2.44& 0.63 & 4.28& 0.12 &0.92 &0.86 & -0.12 &-0.12
      \\
      2nd-order QH & 4 &  $\,$4 & 1.20 & -0.35 & 1.57 & 0.45 & 0.46 &4.56&N/A&-2.30&N/A
      \\
      Precise QA & 2 &  $\,$0 & 1.07 &$\,$2.36 & 0.68 & 3.54 & 0.11 &0.96&0.94&-0.09&-0.09
      \\
      Precise QH & 4 &  -4 & 1.27 & $\,$0.36 & 1.50 & 0.42 & 0.46 & 4.50 &N/A &-2.07&N/A
  \end{tabular}
  \caption{Summary of the  configurations studied in this paper. $N_{\rm fp}$ is the field period. The value of $q_N$ is taken at the axis. The  RH residuals are theoretically calculated at $\epsilon=0.1$. The GAM frequencies and damping rates are normalized to $v_{\rmt\rmi}/R_0$ and are independent from $\epsilon$ since $k_\psi\delta_\psi$ does not depend on $\epsilon$ in our simulations. Here, $\omega^{\rm ana}_\rmr+\rmi\gamma^{\rm ana}$ is the solution of the dispersion function \eqref{eq:GAM_dispersion} and $\omega^{\rm num}_\rmr+\rmi\gamma^{\rm num}$ is obtained from numerical fitting of the simulation data for QA configurations. Numerical fitting is not applicable to QH configurations where GAM oscillations do not exist.}
  \label{table:summary}
  \end{center}
\end{table}
\subsection{Concentric-circular tokamak configurations}
Before presenting the results in QS stellarators, we first show results in several concentric-circular tokamak configurations with different $q$. Although the theoretical and numerical results have been well established for tokamaks, the results shown here will help give an overall picture on the zonal-flow behaviors, in particular the unusual behaviors at small and large $q$. These tokamak configurations can be described analytically in GTC with major radius $R_0=1{\rm m}$ and minor radius at the outer boundary $a=0.1{\rm m}$. The magnetic field is given by $\bd{B}=G_0\nabla\varphi+q^{-1}\nabla\varphi\times\nabla\psi$ with $G_0=B_0R_0$ and $B_0=1{\rm T}$, and the Boozer toroidal angle $\varphi$ is minus the cylindrical toroidal angle, i.e., $\varphi=-\phi$. We simulate 1/24 of the torus with 4 planes, and the other simulation parameters are similar to those described above.

\begin{figure}
    \centering
    \includegraphics[width=1\textwidth]{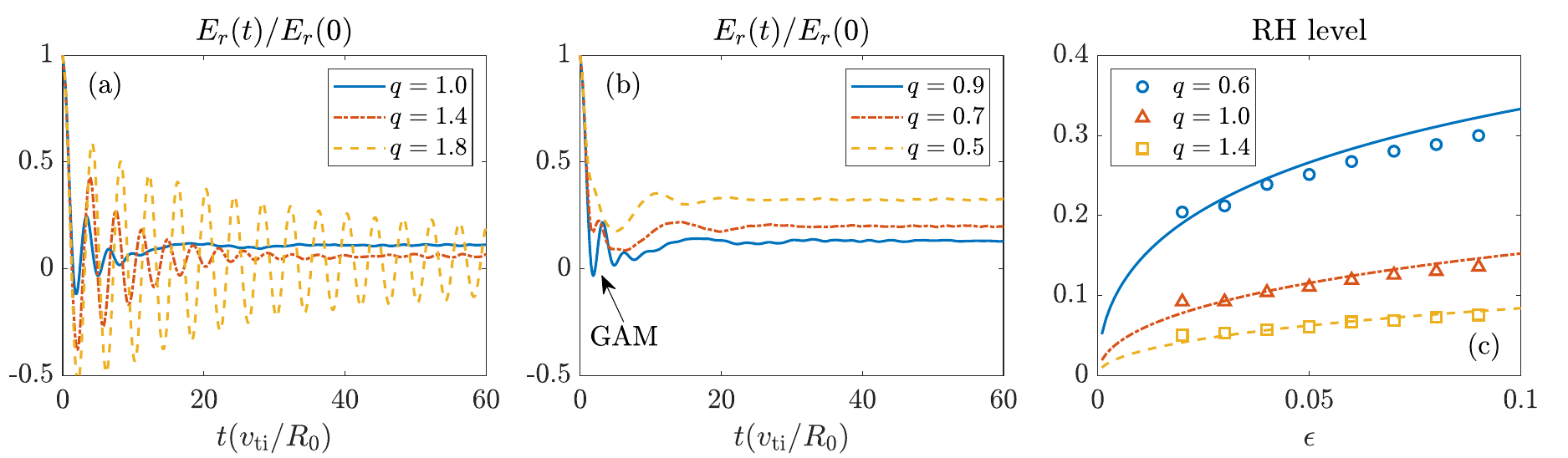}
    \caption{Simulation results for concentric-circular tokamaks. (a): the radial electric field $E_r(t)$  at $\epsilon=0.05$ normalized to its initial value with increasing $q$. At $q=1.8$, GAM oscillations become persistent and do not damp to zero. (b): same as (a) but with decreasing $q$. The GAM oscillations indicated by the text arrow are heavily damped and eventually become non-existent. (c): Comparison between analytical (curves) and numerical (markers) results for the RH level.}
    \label{fig:tokamak}
\end{figure}

Figure \ref{fig:tokamak}(a) shows the results at $q=1.0$, $1.4$, and $1.8$. At $q=1.4$ the zonal flow behaves in the expected way, namely, damped GAM oscillations followed by the RH residual flow. At $q=1.0$ the GAM is quickly damped, followed by a slow relaxation to the RH residual flow. At $q=1.8$, however, GAM oscillations become persistent and do not damp to zero, even though the theory in section \ref{subsec:GAM} still predicts a finite $\gamma<0$. These undamped GAM oscillations occur around $q\approx 1.6$, and they have also been observed in GTC simulations in the past \citep{Lin00} as well as from another global gyrokinetic code COGENT \citep{Dorf13}.

Figure \ref{fig:tokamak}(b) shows the results at $q=0.9$, $0.7$, and $0.5$. As $q$ decreases, the GAM oscillations are heavily damped and eventually become non-existent at $q=0.5$, when the initial perturbation relaxes to the residual flow through a slower oscillation. These slower oscillations cannot be described by the GAM theory in section \ref{subsec:GAM}, since $q<1$ is outside its applicable range.

Finally, figure \ref{fig:tokamak}(c) shows the RH residual level at different $q$, and theory and simulation results  agree well (within a $10\%$ difference). This is expected as the RH flow is a result of the toroidal angular momentum conservation regardless of the GAM behaviors. Also note that the theory and simulation results start to deviate at the smallest $\epsilon$, where the zonal-flow wavelength becomes comparable to the minor radius so that $\epsilon$ itself becomes less well defined.

\subsection{1st-order NAE configurations}
\begin{figure}
    \centering
    \includegraphics[width=1\textwidth]{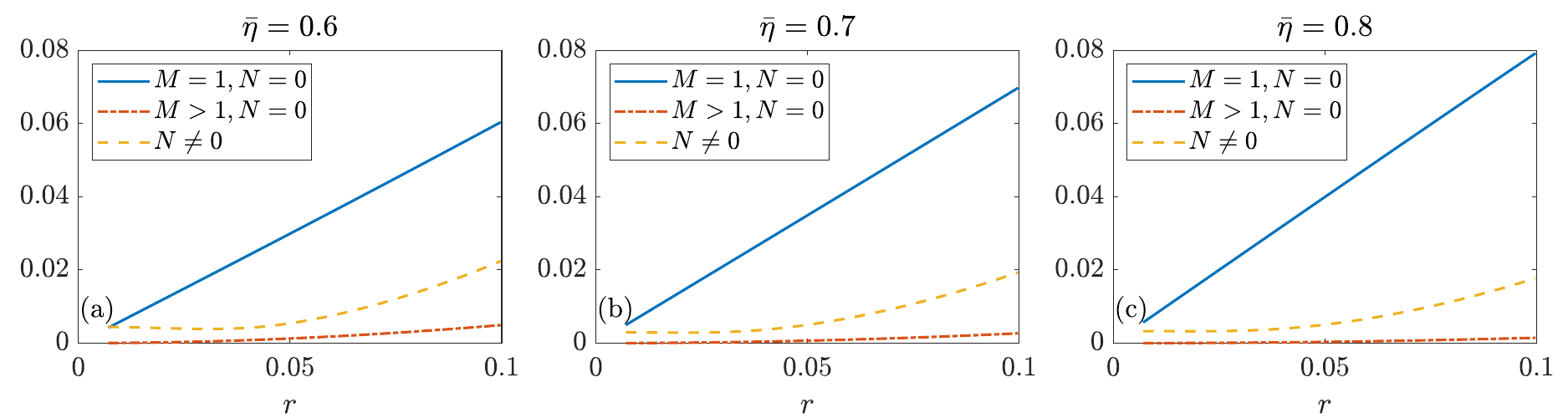}
    \caption{The Fourier spectrum of $B$ for the 1st-order NAE QA configurations. Shown are $\sqrt{\sum B_{MN}^2}$ where the summation is over the range in $(M,N)$ indicated by the legends.}
    \label{fig:NAE_QA1st_error}
\end{figure}

For 1st-order NAE configurations, we follow the examples presented in \cite{Landreman19a}. For QA configurations, the axis shape is chosen to be
\begin{equation}
    \bd{r}_0(\phi)=(1+0.045\cos 3\phi)\bd{e}_R-0.045\sin 3\phi\bd{e}_z,
\end{equation}
where $\phi$ is the cylindrical (not Boozer) toroidal angle. To see the effects from $\etabar$, we compare three different configurations with $\etabar=0.6$, 0.7, and 0.8, which are labeled by ``a'', ``b'', ``c'' in table \ref{table:summary}, respectively. For these configurations, the magnetic-field strength can be written as $B=\sum B_{MN}\cos(M\vartheta-N\varphi)$, where $B_{MN}$ is the Fourier spectrum in Boozer coordinates calculated from BOOZ\_XFORM, and only the cosine components are included due to stellarator symmetry. Figure \ref{fig:NAE_QA1st_error} shows the amplitude of the $N=0$ components, which are QS, and the amplitude of the $N\neq 0$ components, which are QS-breaking. It is seen that $B$ is dominated by the $(M,N)=(1,0)$ QS component, but the $N\neq 0$ QS-breaking components are also significant; in particular, they remain finite near the axis, which seemingly contradicts the NAE description. As mentioned above,  while their boundaries are prescribed by the NAE, these VMEC equilibria are global and not identical to the NAE inside the boundary. \cite{Landreman19b} showed that if we prescribe the boundary at $r=a$  from the 1st-order NAE theory, the axes of the resulting VMEC equilibria will slightly differ from the original axes assumed by the NAE, resulting in a $\mc{O}((a/R_0)^2)$ QS error even at the axes. Therefore, we do not expect these configurations to be close to QS even near the axis.

\begin{figure}
    \centering
    \includegraphics[width=1\textwidth]{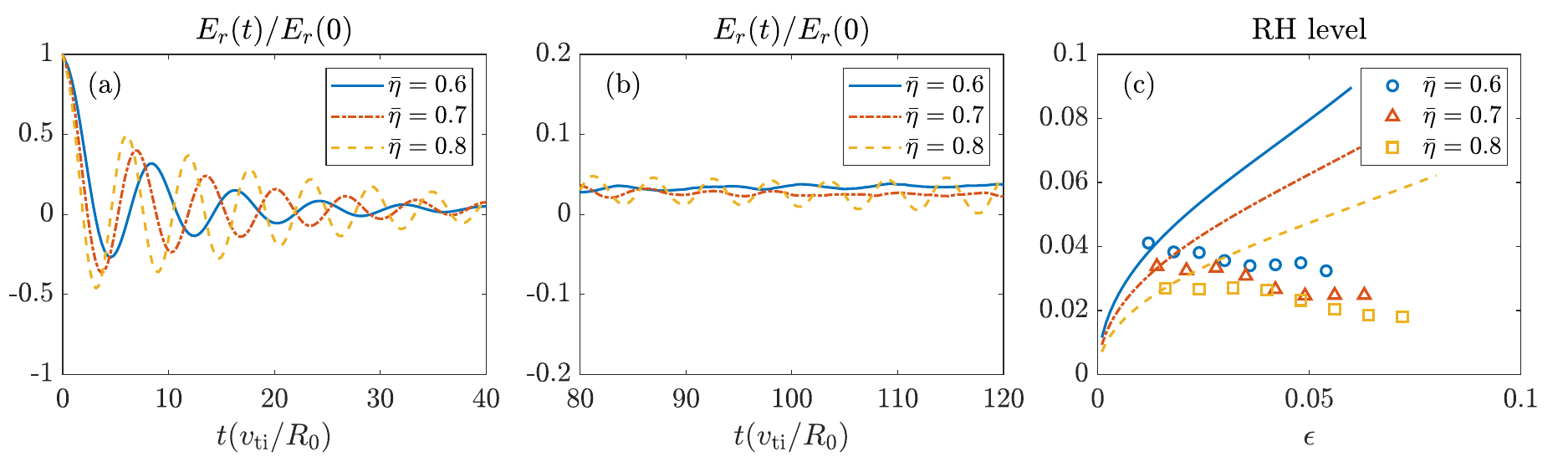}
    \caption{Simulation results for the 1st-order NAE QA configurations. (a) and (b): the radial electric field $E_r(t)$  at $\epsilon=0.05$ normalized to its initial value. The GAM oscillations and the RH levels are shown separately in two figures for a clearer view. (c): Comparison between analytical (curves) and numerical (markers) results for the RH level. The configurations have the same range in $r$ but different range in $\epsilon=\etabar r$ due to their different $\etabar$.}
    \label{fig:NAE_QA1st}
\end{figure}

The GAM oscillations nevertheless behave as expected, which are insensitive to the QS property. As shown in figures \ref{fig:NAE_QA1st}(a) and (b), $\mc{C}$ decreases with increasing $\etabar$ so that the GAM frequency increases. Meanwhile, the GAM damping rate also decreases due to increasing $q_N^2/\mc{C}$. To compare with the analytic results, the simulation results are often fitted with the following formula \citep{Sugama06}: 
\begin{equation}
\label{eq:GAM_fit}
    \frac{E_r(t)}{E_r(0)}={\rm RH}+(1-{\rm RH})\cos(\omega^{\rm num}_\rmr t)\rme^{\gamma^{\rm num} t},
\end{equation}
where RH is the residual level. However, we found it difficult to achieve a globally good fit, because the initial GAM damping rate is much larger than the late-time damping rate as $E_r$ approaches the RH residual. The reason is that as with the typical Landau-damping process, the initial perturbation is not a GAM eigenstate, which only emerges at large $t$ after the initial fast damping due to phase mixing. Therefore, we ignore the initially large GAM damping rates, and numerically find $(\omega^{\rm num}_\rmr, \gamma^{\rm num})$ that matches $E_r$ as it approaches the RH level. Comparison with solutions of the dispersion function \eqref{eq:GAM_dispersion} are shown in table \ref{table:summary}, and both $\omega_\rmr$ and $\gamma$ agree well with the theoretical prediction. Also note that the GAM oscillations do not completely damp to zero at $\etabar=0.8$ where $|q_N|/\sqrt{C}=1.65$, consistent with the observation in figure \ref{fig:tokamak}.


For the RH level, however, numerical results do not agree with the theoretical predictions. As shown in figure \ref{fig:NAE_QA1st}(c), theory and simulation results do not show any agreement. Further, as $\epsilon$ increases, the numerical RH level actually decreases, in contrast to the theory. This is not a surprise considering the large QS breaking components shown in figure \ref{fig:NAE_QA1st_error}. In fact, \citet{Helander11} studied the effects of radially unconfined trapped particles and  found the long-time residual level to be
\begin{equation}
       \frac{E_r(\infty)}{E_r(0)}=\left[1+\frac{\alpha q^2}{\sqrt{\epsilon}}+\frac{\beta\sqrt{\epsilon}}{(k_\psi\rho_\psi)^2}\right]^{-1},
\end{equation}
where the factor $\beta$ comes from the unconfined particles.  At small $|k_\psi\rho_\psi|$, $\beta/(k_\psi\rho_\psi)^2$ can be large and hence can provide a possible explanation for the observed decrease in the RH level at large $\epsilon$. 

For the 1st-order QH configurations, the example presented in \cite{Landreman19b} has $a=0.025{\rm m}$, so it is 4 times thinner than the 1st-order QA configurations. The reason is that due to the strongly shaped axis, the 1st-order QH configuration achieves the same level of QS-breaking components in $B$ at a 4 times smaller $r$ compared to the 1st-order QA configuration. Therefore, we expect even more significant QS errors for the 1st-order QH configuration at larger radius $a\sim 0.1{\rm m}$, so we skip this configuration and proceed to 2nd-order NAE configurations below.

\subsection{2nd-order NAE configurations}
\begin{figure}
    \centering
    \includegraphics[width=0.67\textwidth]{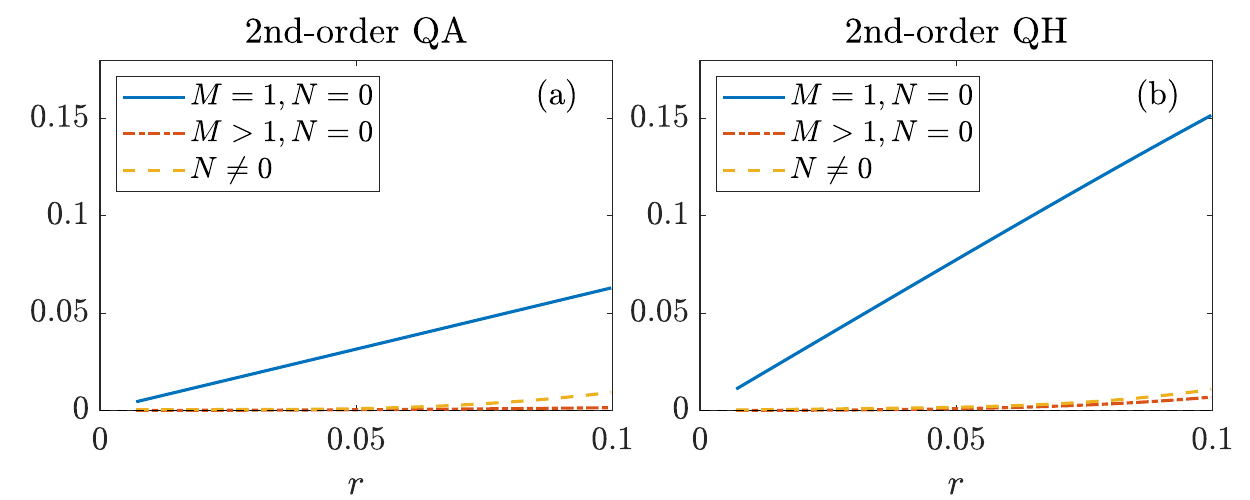}
    \caption{The Fourier spectrum of $B$ in helical angle $(\vartheta,\varphi)$ for the 2st-order NAE QA and QH configurations. Shown are $\sqrt{\sum B_{MN}^2}$ where the summation is over the range in $(M,N)$ indicated by the legends.}
    \label{fig:NAE_2nd_error}
\end{figure}
\begin{figure}
    \centering
    \includegraphics[width=1\textwidth]{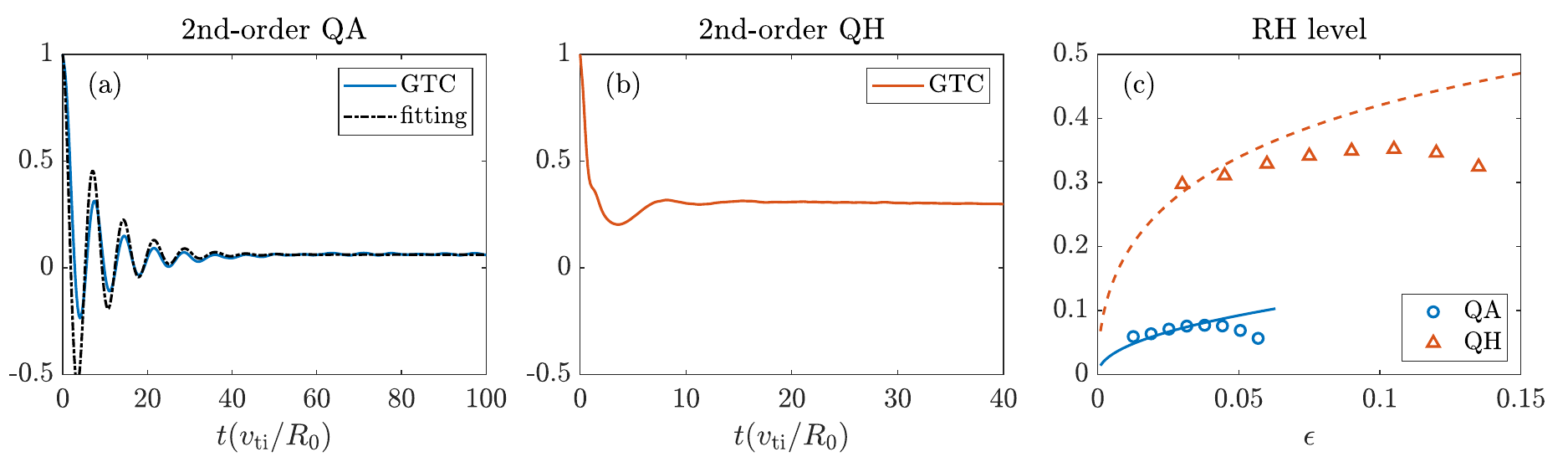}
    \caption{Simulation results for the 2nd-order NAE QA and QH configurations. (a) and (b): $E_r(t)$ at $\epsilon=0.05$ for QA and QH. The black dashed curve is from the numerical fit \eqref{eq:GAM_fit}. (b): Comparison between analytical (curves) and numerical (markers) results for the RH level. The QH configuration has the same range in $r$ as the QA configuration, but a larger range in $\epsilon=\etabar r$ due to its larger $\etabar$.}
    \label{fig:NAE_2nd}
\end{figure}

We have seen that for the 1st-order NAE configurations, the QS-breaking components of $B$ are significant, resulting in disagreement in theory and simulation result on the RH level. To see if such deviation can be reduced with reduced QS error, we test the 2nd-order NAE QA and QH configurations from \citet{Landreman19b}. For the 2nd-order QA configuration, the axis is chosen to be
\begin{multline}
    \bd{r}_0(\phi)=\left(1+0.173\cos2\phi+0.0168\cos 4\phi+0.00101\cos6\phi\right)\bd{e}_R\\
    +\left(0.159\sin2\phi+0.0165\sin4\phi+0.000985\sin6\phi\right)\bd{e}_z,
\end{multline}
with $\etabar=0.632$. For the 2nd-order QH configuration, the axis is
\begin{multline}
    \bd{r}_0(\phi)=\left(1+0.17\cos4\phi+0.01804\cos 8\phi+0.001409\cos12\phi+0.00005877\cos16\phi\right)\bd{e}_R\\
    +\left(0.1583\sin4\phi+0.0182\sin8\phi+0.001548\sin12\phi+0.00007772\sin16\phi\right)\bd{e}_z,
\end{multline}
with $\etabar=1.569$. The normal vector $\bd{n}$ rotates around the axis poloidally four times as the axis is traversed toroidally, resulting in $N=4$. For these configurations, the boundary at $r=a$ are carefully chosen so that the axes of the resulting VMEC equilibria are much closer to the original axes assumed by the NAE, which reduces the QS error at the axis to $\mc{O}((a/R_0)^3)$. As shown in figure \ref{fig:NAE_2nd_error}, the QS-breaking components of $B$ are much smaller compared to the 1st-order configurations near the axis (figure \ref{fig:NAE_QA1st_error}). However,  a toroidal variation in $B$ has to be introduced in order to construct these configurations, which is zero at the axis and increases with $r$ as $\mc{O}(r/R_0)^2$. Therefore, strictly speaking, the QS-breaking components remain at $\mc{O}(\epsilon^2)$ rather than $\mc{O}(\epsilon^3)$ for the 2nd-order NAE configurations.

Numerical results are shown in figure \ref{fig:NAE_2nd}. For the 2nd-order QA configuration, the GAM oscillations are very similar to the 1st-order QA in figure \ref{fig:NAE_QA1st}; the numerical fitting formula \eqref{eq:GAM_fit} provides a reasonable description at large $t$, and the numerical and theoretical frequencies also agree. Meanwhile, the RH residual agrees much better with theory at small $\epsilon$, but still deviate from theory at large $\epsilon$ due to the increasing QS error. For the 2nd-order QH configuration, $q_N/\sqrt{\mc{C}}=0.48$ and $E_r$ quickly drops to the RH residual without GAM oscillations, consistent with the result in tokamaks with $q=0.5$ (figure \ref{fig:tokamak}) as well as previous numerical results from simulations of zonal flows in HSX \citep{Smoniewski21}. Since GAM oscillations do not exist, the numerical fitting \eqref{eq:GAM_fit} are not applicable to the QH configuration. Also, despite the small $\mc{C}$, the RH level in the QH configuration is still much larger than the QA configuration due to the small $|q_N|$, as predicted by an earlier study \citep{Plunk24}. However, the simulated RH level is still much lower than the theoretical prediction, indicating that the QS-breaking components of $B$ are still significant.

\subsection{The precise QA and QH configurations}
\label{sec:simulation_precise}
\begin{figure}
    \centering
    \includegraphics[width=0.67\textwidth]{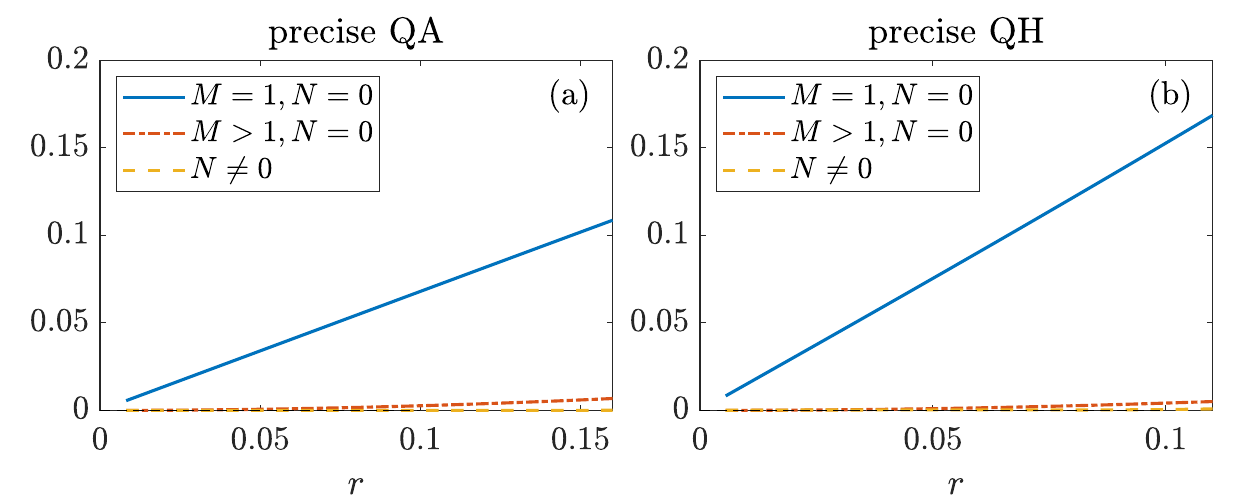}
    \caption{The Fourier spectrum of $B$ in helical angle $(\vartheta,\varphi)$ for the precise QA and QH configurations. Shown are $\sqrt{\sum B_{MN}^2}$ where the summation is over the range in $(M,N)$ indicated by the legends.}
    \label{fig:precise_error}
\end{figure}
\begin{figure}
    \centering   \includegraphics[width=1\textwidth]{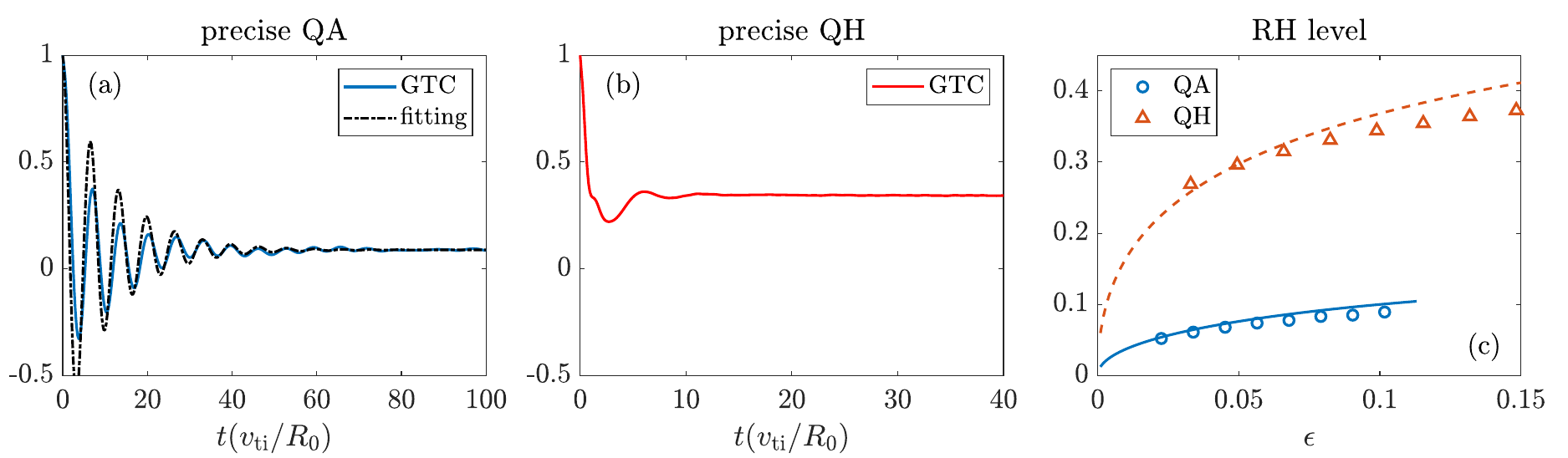}
    \caption{Simulation results for the precise QA and QH configurations. (a) and (b): $E_r(t)$ at $\epsilon=0.1$. The black dashed curve is from the numerical fit \eqref{eq:GAM_fit}. (b): Comparison between analytical (curves) and numerical (markers) results for the RH level. The precise QH configuration has a smaller range in $r$ than the precise QA configuration, but still a larger range in $\epsilon=\etabar r$ due to its larger $\etabar$.}
    \label{fig:precise}
\end{figure}

The precise QA and QH configurations are obtained from global optimization using the software framework SIMSOPT \citep{Simsopt}. As shown in figure \ref{fig:precise_error}, the QS-breaking components of $B$ are very close to zero. Also, the QS components of $B$ are still dominated by $M=1$, so that $B=B_0[1+\epsilon\cos\vartheta+\mc{O}(\epsilon^2)]$ holds even though they are not generated from the NAE approach. Quantities such as $\etabar$ and $\sigma$ can also be obtained near the axis and used to calculate $\mc{C}$, which showed good agreement with direct numerical evaluation of $\avg{|\nabla\psi|^2/B^2}$. As shown in figure~\ref{fig:precise}, numerical results of the GAM dynamics are qualitatively similar to the NAE configurations. For the RH residual, good agreement between theory and numerical results can be achieved throughout the volume for both the QA and QH configurations, the difference being less than 10\%. Therefore, the theoretical description of collisionless zonal-flow dynamics can be applicable to actual QS stellarator configurations when the QS-breaking components of $B$ become small enough.

\section{Conclusions}
\label{sec:conclusion}
The linear collisionless plasma response to a zonal density perturbation in QS stellarators is studied, including the GAM oscillations  and the RH residual-flow level.  It is found that while the GAM oscillations in QA configurations are similar to tokamaks, they become non-existent in QH configurations due to the small effective safety factor $q_N$ in helical-angle coordinates. Compared with concentric circular tokamaks, the RH residual is also found to be modified by a geometric factor $\mc{C}$, which we derived analytically using the NAE framework. It is found that $\mc{C}>1$ for the QA configurations and $\mc{C}<1$ for the QH configurations studied in the paper. Nevertheless, the QH configurations still have much larger RH residual due to the much smaller $q_N$. These analytic results are compared with numerical simulation results from  GTC.  While the GAM physics is reasonably predicted by the theory, we found that for the RH residual level, good agreement between analytical and numerical results is achieved only when the amplitude of QS-breaking magnetic-field component is small enough. Since zonal flows can be important for regulating turbulent transport, these results suggest possible relation between the transport level and the stellarator geometric parameters via nonlinear interactions with zonal flows.

The data that supports the findings of this study are openly available at Zenodo \citep{Zhu24data}.

H.Z. thanks W. Sengupta, R. Jorge, E. Rodr\'{i}guez, E. Green, X. Wei for useful discussions. H.Z. was supported by a grant from the Simons Foundation/SFARI (Grant \#560651, AB). This research used resources of the National Energy Research Scientific Computing Center, which is supported by the Office of Science of the U.S. Department of Energy under Contract No. DE-AC02-05CH11231, and the Department of Energy  SciDAC HiFiStell project supported by Contract No. DE-SC0024548. 

\bibliographystyle{jpp}

\bibliography{refs}

\end{document}